\documentclass[12pt,preprint]{aastex} 		% single spaced
\newcommand{\kms}{km s$^{-1}$}
\newcommand{\solarmass}{\ensuremath{ \mathnormal{M}_{\Sun} }}
\newcommand{\mass}{\ensuremath{ \mathnormal{M} }}

\received{2005 June 2}
\begin{document}

\title{Spitzer Observations of G Dwarfs in the Pleiades: Circumstellar Debris
Disks at 100 Myr Age\altaffilmark{1,2}}

\slugcomment{To Appear in the Astronomical Journal} 
\shortauthors{Stauffer et al.} 
\shorttitle{G dwarfs in the Pleiades}

\author{John R. Stauffer}
\affil{Spitzer Science Center, Caltech 314-6, Pasadena, CA  91125}
\email{stauffer@ipac.caltech.edu}

\author{Luisa M. Rebull} 
\affil{Spitzer Science Center, Caltech 220-6, Pasadena, CA  91125}

\author{John Carpenter and Lynne Hillenbrand}
\affil{Astronomy Department, California Institute of Technology,
Pasadena, CA  91125 }

\author{Dana Backman} 
\affil{SOFIA / SETI Institute, MS 211-3, NASA - Ames Research Center, 
Mountain View, CA  94035-1000}

\author{Michael Meyer, Jinyoung Serena Kim, Murray Silverstone and Erick Young}
\affil{Steward Observatory, University of Arizona, Tucson,
AZ  85726}

\author{Dean C. Hines}
\affil{Space Science Institute, 4750 Walnut Street, Suite 205 Boulder, CO 80301}

\author{David R. Soderblom}
\affil{Space Telescope Science Institute, 3700 San Martin Dr.,
Baltimore, MD  21218}

\author{Eric Mamajek}
\affil{Center for Astrophysics, 60 Garden St., Cambridge, MA 02138}

\author{Patrick Morris} 
\affil{Infrared Processing and Analysis Center, Caltech 314-6, Pasadena, CA  91125}

\author{Jeroen Bouwman}
\affil{Max-Planck Institut fur Astronomie, Heidelberg, Germany}

\author{Stephen E. Strom}
\affil{National Optical Astronomy Observatory, 950 N. Cherry Avenue,
Tucson, AZ  85719}

\altaffiltext{1}{This work is based (in part) on observations made
with the Spitzer Space Telescope, which is operated by the Jet 
Propulsion Laboratory, California Institute of Technology, under
NASA contract 1407. } 
\altaffiltext{2}{This publication makes use of data products 
from the Two Micron All Sky Survey, which is a joint project 
of the University of Massachusetts and the Infrared Processing 
and Analysis Center/California Institute of Technology, 
funded by the National Aeronautics and Space Administration 
and the National Science Foundation.}

\begin{abstract}

Fluxes and upper limits in the wavelength range from 3.6 to 70 $\mu$m\
from the Spitzer Space Telescope are provided for twenty solar-mass
Pleiades members.  One of these stars shows a probable mid-IR
excess and two others have possible excesses,
presumably due to circumstellar debris disks.   For the
star with the largest, most secure excess flux at MIPS wavelengths, HII1101,
we derive Log(L$_{dust}$/L$_*$) $\sim$ -3.8 and
an estimated debris disk mass of 4.2$\times$10$^{-5}$\ M(Earth) for
an assumed uniform dust grain size of 10$\mu$m.
If the stars with detected
excesses are interpreted as stars with relatively recent, large
collision events producing a transient excess of small dust particles,
the frequency of such disk transients is about $\sim$10$\%$ for our $\sim$100
Myr, Pleiades G dwarf sample.
For the stars without detected 24-70$\mu$m excesses, the upper limits
to their fluxes correspond to approximate 3$\sigma$\ upper limits
to their disk masses of 6$\times$10$^{-6}$\ M(Earth) using the MIPS
24$\mu$m upper limit, or 2$\times$10$^{-4}$\ M(Earth) using the
MIPS 70$\mu$m limit.   These upper limit disk masses
(for ``warm" and ``cold" dust, respectively)
are roughly consistent, but somewhat lower than, predictions of a
heuristic model for the evolution of an ``average" solar-mass star's
debris disk based on extrapolation backwards in time from current
properties of the Sun's Kuiper belt.

\end{abstract}


\section{Introduction}
\label{sec:intro}

The ``Formation and Evolution of Planetary Systems" (FEPS) Spitzer
Legacy program is observing a sample of 328 relatively nearby low mass
stars with masses in the range 0.8 $<$ \mass/\solarmass $<$ 1.5 and with a range
of ages from 3 Myr to about 3 Gyr \citep{meyer04}.  The goal of the program is to
determine empirically the ubiquity and evolution of circumstellar dust
disks around approximately solar-mass stars, and hence to put
constraints on the frequency and timescale for the formation of planets
around such stars.   All of the program stars are being observed with
all three of Spitzer's instruments: IRAC (3.6, 4.5 and 8.0 microns) in order
to help define the photospheric flux at short wavelengths and to look for
``hot" dust analogous to the Solar System's inner zodiacal dust; IRS
(5 to 38 micron low resolution spectra) in order 
to place constraints on the particle size and mineralogy; and
MIPS (24 and 70 microns, and in some cases 160 microns) in order to
place limits on the presence of cool 
dust around these stars.

Most of the FEPS sample of stars are field dwarfs whose ages are estimated
from a variety of somewhat indirect techniques.  However, 
the FEPS sample also includes about
120 stars in several open clusters or associations, chosen to sample a range
in age from a few Myr to about 600 Myr.  These stars have the advantage that
they have much better determined ages than the field stars; however, they
are also on average considerably further away than the other stars in the
FEPS sample, and therefore the ability to detect disks with
Spitzer data is diminished.
The clusters included in the FEPS target list are:  the Hyades (age $\sim$
600 Myr); the Pleiades \citep[age $\sim$ 100 Myr,][]{meynet93}; 
Alpha Persei \citep[age $\sim$ 75 Myr,][]{ventura98,stauffer99}); 
IC2602 \citep[age $\sim$ 45 Myr,][]{merm81,stauffer97,barrado04}; 
and Sco Cen \citep[age $\sim$ 5-17 Myr,][]{preibisch01,mamajek02}.   
In this paper, we will
describe the Spitzer observations we have obtained of the Pleiades stars
that are included in the FEPS program.   
Note that a Spitzer guaranteed time program led by J. Muzerolle has 
obtained Spitzer data for higher mass stars in the Pleiades. 
Those data will be reported in a separate paper.

Observations of the low mass
stars in young open clusters show that those stars arrive on the main
sequence with a wide range of rotational velocities.  Half or more 
arrive on the main sequence with low rotational velocities
(vsini $<$ 20 \kms); the remainder arrive on the main sequence as
rapid rotators, with rotational velocities up to almost 200 \kms.
A commonly proposed model for explaining the wide spread in ZAMS
rotational velocity is that angular momentum is transferred to 
circumstellar disks during PMS evolution via magnetic fields (``disk
locking"), and hence the stars that are slowly rotating upon arrival
on the ZAMS are those that had the longest-lived PMS accretion disks
\citep{bouvier93}.  If one
further assumes that stars with long-lived accretion disks also are
likely to have longer-lived or more massive debris disks, then there
should be a correlation between rotational velocity and mid-IR 
excess for young, low mass stars.  An age of order 100 Myr (i.e.
the Pleiades age) should be
ideal for looking for this correlation, because main-sequence angular
momentum loss from winds will not yet have significantly affected the
rotational velocities and the debris disks should be relatively massive
and bright in the mid-IR compared to such disks surrounding older
main sequence stars.  The Pleiades low mass stars 
exhibit the large rotational velocity range that is desired for this
experiment \citep{vleeuwen86,stau87,queloz98};
Spitzer may be able to provide the mid-IR data necessary to link
these rotational velocities to disk evolution in the terrestrial
and jovian planetary formation zones.

There are two other correlations between physical properties of the
parent star and detectability of debris disks that have been reported
in recent papers.  \citet{chen05} obtained MIPS observations
of a sample of about 40 F and G type members of the Sco-Cen association
(age $\sim$ 5-20 Myr) and detected about one third of them at 24$\mu$m.
They found an anti-correlation between disk excess and xray activity
for their sample. They interpreted this in terms of the xray bright
stars having strong winds, which could scour small dust
particles from the debris disks in these systems.  However, another correlation
present in the \citet{chen05} data was that the stars with detected
debris disks were preferentially earlier spectral types in
their sample (hence higher mass if approximately the same age).  A
similar correlation between stellar mass and debris disk detection was noted
by \citet{wyatt03} in a survey of the \citet{lindroos86} sample of
young, nearby stars.   The correlation between
disk detection frequency and spectral type or mass may simply arise
because smaller amounts of dust are  more easily detected around more 
luminous stars, and therefore the correlation between disk excess and
xray activity found by \citet{chen05} may not require winds as the
physical explanation.   Determining if there is a correlation between
disk excess and xray activity for a sample of stars with a wide range
of xray activity but a small range of mass would be a valuable step.

There have been several previous discussions of and/or searches for 
circumstellar dust around low mass Pleiades members.  \citet{jones72}
suggested that K dwarfs in the Pleiades were surrounded by dust shells
in order to explain why those stars fell below the main sequence in a
M$_V$ vs. B-V CMD \citep[for an alternative explanation of the
CMD behavior not involving dust shells, see][]{stauffer03}.
\citet{back91} used the IRAS 25$\mu$m and
60 $\mu$m data to search for IR excesses around a sample of Pleiades and 
IC2391 A and early F dwarfs.  They reported finding weak evidence for detecting
an excess for the ensemble of IC2391 stars, but did not detect an excess
for the Pleiades sample.  \citet{meyer00} used data from ISO for another
sample of A and F dwarfs in several nearby, young open clusters, including
the Pleiades, but again
were unable to detect IR excesses unambiguously in these stars. 
\citet{spangler01} also used the ISOPHOT instrument on ISO to search
for debris disks around young stars.
They observed 14 Pleiades low mass members at 60 and 90 $\mu$m, with
possible detections at one or both wavelengths for two of those
members (HII1132, an F dwarf, and HII3163, a K dwarf).
MSX was also used to observe the Pleiades \citep{kraemer03}, but only 
detected the very brightest (B star) members of the cluster longwards 
of 10$\mu$m.

\section{Properties of the Pleiades Stars Observed with Spitzer}
\label{sec:sample selection}

All of the main sequence stars in the FEPS program were required
to have 0.5 $<$ B-V $<$ 1.0, and that criterion was applied to the
Pleiades sample.   Another significant constraint on our selection
of Pleiades stars was that Legacy observations were required not
to conflict with GTO targets, and the Muzerolle et al. 
GTO program obtained a one square degree map of the center of 
the Pleiades using both IRAC and MIPS.  Our FEPS targets are therefore
located somewhat on the outskirts of the cluster.   
We also attempted to minimize difficulties caused by the
extended IR emission from the gas and dust associated with the CO
cloud impacting the Pleiades \citep{white03} by avoiding stars
within the region of high extinction identified by \citet{breger87}.
Finally, some preference in selecting FEPS targets
was given to stars not known to be binary.  With these 
constraints in mind, the set of Pleiades members selected for
Spitzer observation is provided in Table 1.  All of these stars
are very high probability members of the cluster based on extensive
radial velocity and proper motion studies, and on other indirect
measures such as chromospheric and coronal activity and lithium
abundance \citep{stau87,soderblom93}.

The sources of the data in Table 1 are described in detail in
\citet{carpenter05}; however, we provide a brief synopsis here.
The V and B-V data are generally derived from averaging the 
observations reported in \citet{jomi58}, \citet{mendoza67}, \citet{stau84},
and \citet{stau87}.  The spectroscopic rotational velocities
are from \citet{queloz98}, except for HII174 \citep[from][]{wh05}
and HII1101 \citep[from an average of values in][]{stau87,merm93}.
Rotation periods have been reported for HII 250 \citep{mess01,marilli97};
HII314 \citep{prosser93}; HII2786 \citep{krishna98}; 
and HII2881 \citep{prosser93}.  
The rotation period for HII250 may be spurious given the very
small vsini for this star - the star would have to be nearly pole-on if
both the vsini and rotation period are correct,
but in that case one would expect very little photometric rotational
modulation.  Finally, the following stars in our list have been
identified as binaries:
HII120 \citep{raboud98}; HII173 \citep{shs84,merm92};
HII1182, HII2278 and HII2881 \citep{bouvier97}; 
HII2147 \citep{queloz98}; and HII3097 \citep{merm92}.
HII1101 was also reported as being a spectroscopic binary in
\citet{soderblom93} and \citet{merm92}.  However, we believe that
identification was spurious - careful examination of a number of
spectra of this star does not support the SB designation.  The
xray data [Log(L$_X$/L$_{Bol}$)]in Table 1 are from \citet{stauffer94}
and \citet{micela99}.

Compared to the rest of the FEPS sample, the Pleiades sample might
appear to include a higher fraction of binaries.  However, that is
at least in part illusory, owing to the fact that 
the Pleiades have been searched for binary stars more thoroughly
than most of the FEPS targets.  For
example, \citet{merm92} and \citet{raboud98} obtained more than
ten years of accurate radial velocity monitoring of most of the
Pleiades F, G and early K dwarfs, allowing very good detection of short and
intermediate period systems.  \citet{bouvier97} obtained JHK AO imaging
with CFHT in 1996 of most of the same stars, allowing detection of long
period binary companions with separations of 0.1 to 6.9 arcsec and
and K-band flux ratios up to 100. \citet{metchev05} obtained K
band AO imaging of the Pleiades FEPS sample in 2002, allowing
determination of common proper motion for possible companions detected
in both surveys.  Three of the Pleiades binaries would probably not
have been identified as such if data similar to what is available for
the average FEPS field dwarf were utilized.  For example, HII3097, 
an SB1 with P=774 days, e = 0.78 and K1 = 5.1 \kms\ \citep{merm92} 
has radial velocities which remain within about
2 \kms\ over about 80\% of its orbital phase due
to the high eccentricity of its orbit.
Similarly, HII120 is listed as a possible SB1 by \citet{raboud98},
but with only a 3 \kms\ range in radial velocity over a 13
year timespan; this star would also have been unlikely to have been
flagged as a binary in the rest of the FEPS sample.  The companion
to HII1182 \citep{bouvier97} was also detected in the FEPS-team AO
imaging program, but with $\Delta$K $\sim$ 4.8 and separation = 1.14
arcseconds, we could have only listed it as a possible companion
pending determination of common proper motion.   Therefore, of the
seven probable binaries in the Pleiades sample, only four would
likely to have been identified as such if they had been in our field
star sample.  The characteristics of the four ``easily detected"
binary systems in our Pleiades sample are:
\vskip0.1truein
\noindent
HII173:  SB2, P = 497 days, q $\simeq$ 0.95 \citep{merm92}
\vskip0.1truein
\noindent
HII2147: SB2, P $>$ 15 years, q$\sim$ 0.82 \citep{queloz98,merm92}
\vskip0.1truein
\noindent
HII2278: VB, $\Delta$K = 0.05, separation 0.37 arcsec \citep{bouvier97}, q
$\sim$ 1.00 \citep{merm92}
\vskip0.1truein
\noindent
HII2881: VB, $\Delta$K $\sim$ 0, separation 0.08 arcsec \citep{bouvier97}, q
$\sim$ 0.81 \citep{merm92}

\begin{figure} %
     \includegraphics[angle=00,totalheight=6.5in]{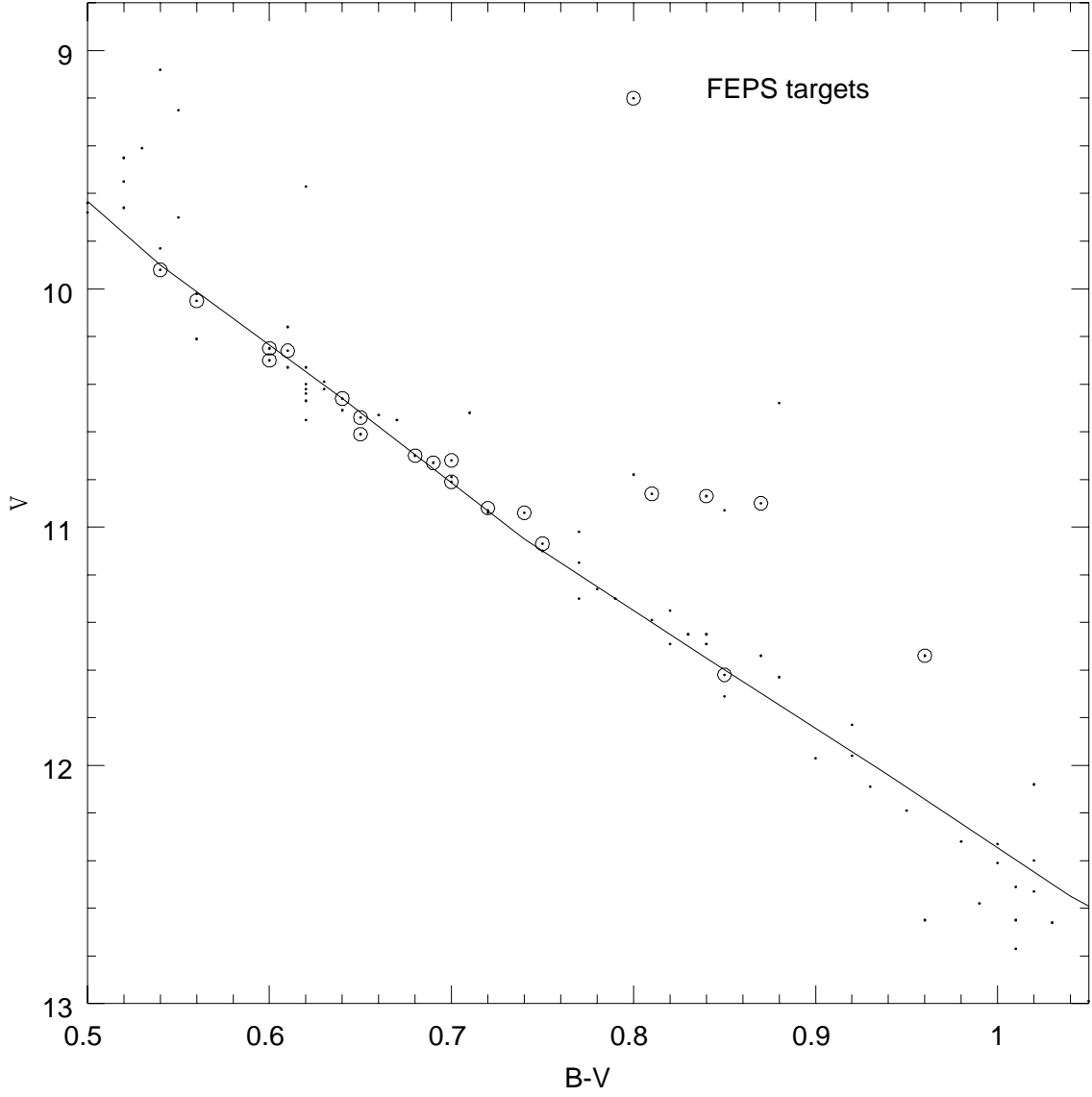}
    \caption{
	$V$ versus $B-V$ color-magnitude diagram for Pleiades members derived
        from photoelectric photometry tabulated in \citet{carpenter05}.  The
        FEPS targets are marked as circled dots.
    \label{fig:plecmd}
    }
\end{figure}

\begin{figure} %
     \includegraphics[angle=00,totalheight=6.5in]{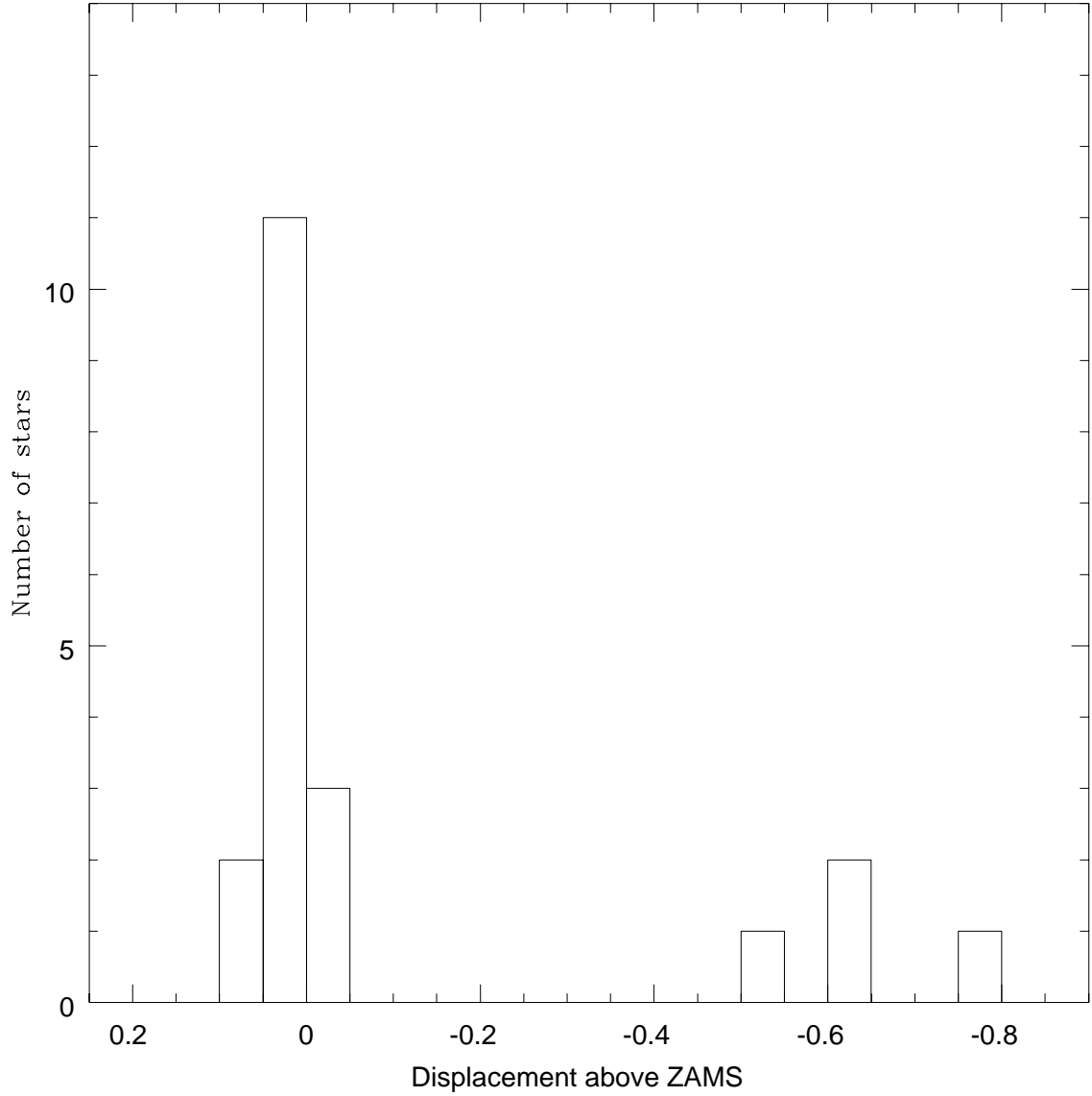}
    \caption{
       Displacement above the ZAMS curve of Figure 1 for the Pleiades FEPS
       sample.  The four stars to the right hand side of the diagram are all
       known binaries with inferred secondary to primary mass ratios near one.
    \label{fig:plehist}
    }
\end{figure}

Because the Pleiades stars are members of a
cluster, they have a very well determined age, distance, and 
reddening, and hence a color-magnitude diagram can quite accurately
identify binaries if the two stars have similar masses.  We
illustrate the latter point here as our Figure \ref{fig:plecmd}.
The FEPS stars are indicated by circled dots, and the solid line
is simply an empirical ZAMS (drawn to follow the single-star main
sequence defined by all Pleiades members - \citet{johnson58}).  
The ``easily detected" binary systems are simple to
identify in the plot - they are the four FEPS stars displaced above the
main sequence curve by more than 0.5 mag; their displacement in the 
CMD is as expected from the fact that they are either SB2's
or visual binaries with nearly equal magnitude components.
Figure \ref{fig:plehist} provides a histogram of the displacements
of the Pleiades FEPS stars relative to the ZAMS curve, and indicates
that with the exception of the four q $\sim$ 1 binaries, the other
FEPS Pleiades stars all lie within 0.1 mag of the single star locus.
Based solely on the CMD, therefore, one can conclude that these
sixteen stars do not have companions within about half the mass
of the ``primary" (at any separation), or, correspondingly, do not
have companions within about 4 magnitudes as bright as the ``primary"
\citep{merm92}.

\section{Spitzer Observations} 
\label{sec:observations}

The IRAC data for 19 of the 20 Pleiades FEPS targets were obtained 
during a single IRAC campaign in early September 2004 (the one
exception is HII1776, for which IRAC observations had not been
obtained at the time of the writing of this paper).  All of the
IRAC observations were obtained in sub-array mode, with data being
obtained in channels 1, 2 and 4 (with central wavelengths of 3.6,
4.5 and 8.0 microns).  The integration time per exposure was set to 0.08
seconds for the three brightest stars, and to 0.32 seconds for the
remainder.  Each target was observed at four dither positions.  In
subarray mode, 64 exposures are obtained at each position,
and therefore 256 exposures were obtained for
each star.  The pipeline processing software used at the Spitzer
Science Center (SSC) for the data we analysed was S10.5.  
\footnote{The IRAC Channel 1 flux calibration in the S10.5 pipeline had
a slight systematic error (which was corrected in S11).  We have adjusted
our derived in-band fluxes for channel 1 downward by 7\% to correct for this,
consistent with the improved S11 flux calibration.}
We used the analysis package IDP3 \citep{schneider02} 
to derive aperture photometry from the BCD (Basic Calibrated
Data) images provided to us by the SSC.
An aperture size of 3 pixels was used,
with multiplicative aperture corrections of 1.112, 1.113 and 1.218,
respectively, for the three channels.  See \citet{silverstone05}
for more details concerning the method used to analyze the IRAC BCD images.

IRS spectra for all 20 Pleiades FEPS targets were obtained
during September 26 and 27, 2004.   The spectra cover the wavelength range
7.4 to 38 $\mu$m at relatively low resolution ($\lambda$/$\Delta$$\lambda$ $\simeq$
70 - 120).  We began each observation (each ``AOR") with a ``high accuracy" peakup
using the short-wavelength side of the IRS peak-up array, in order to
insure that the Pleiades star was accurately centered in the spectrograph
slits.   This worked well except in one case - for HII 2881, another
object (presumably an IR bright, distant galaxy) was brighter than HII 2881
in the peakup array bandpass, and the spectra we obtained were therefore
of the galaxy.  For this reason, we do not have IRS spectra of HII 2881.  The
total integration times for our targets depended on the predicted
mid-IR flux, with a minimum of about 90 seconds in the short-low module
and 500 seconds in the long-low modules to a maximum of about 240 seconds
in the short-low module and 2770 seconds in the long-low modules.  
As the starting point for our analysis, we used the 2-D 'droop' spectral 
images i.e.  the intermediate
data product produced by the S11.0.2 pipeline, prior to
flat-fielding or stray-light correction. To extract the spectra,
we use the SMART software package
(Higdon et al. 2005). The spectra where extracted using a fixed-width
aperture of 6.2 pixels for the first order of the
short wavelength low-resolution module and 5.1 and 3.1 pixels for the
first and second order of the long wavelength low-resolution
module, respectively. We corrected for the sky background and posible
straylight by subtracting the two nod positions observed along
the slit. The spectra where calibrated using a spectral response function 
derived from a set IRS spectra and stellar models of 16 stars observed 
within the FEPS program (see Bouwman et al. 2005, and
the explanatory suplement to the FEPS data at the SSC data archive for
further details on IRS data reduction).
For the purpose of using the spectra to test quantitatively for IR
excesses above the photospheric flux, we have defined pseudo-bandpasses at
24 and 33 microns, assuming a filter response function for the IRS24 band
identical to that for MIPS24 and a square bandpass running from 30 to 35
microns for IRS33.  We integrated the spectra over these
bandpasses to derive in-band fluxes which we designate as IRS24 and IRS33.

The MIPS data for 19 of the 20 Pleiades FEPS targets were obtained
during a single MIPS campaign - more specifically, the Pleiades
stars were observed in an eight hour window on September 22, 2004
(with all 19 targets being observed consecutively).  The one
star lacking MIPS data is again HII1776.  MIPS small-field
photometry mode was used.
At 24$\mu$m, four cycles of 10 second exposures were 
obtained for all of the targets, corresponding to a total of 
about 600 seconds of integration time per source.   At 70$\mu$m, the on source
time was varied depending on the brightness of the star and the expected
background, with a range from about 230 seconds to about 1090 seconds.
For 24$\mu$m, we based our analysis on BCD and post-BCD data from the
S10.5 pipeline processing by the SSC.
Because the Pleiades stars are relatively faint and the
``background" (Pleiades nebular structures, field stars and background
AGN) is complex, we deviated from the standard FEPS analysis of
the MIPS 24$\mu$m  data.   In particular, instead
of aperture photometry we derived PSF-fitting photometry using the
APEX software package \citep{makovoz05} with a 24$\mu$m PSF constructed
from a MIPS map obtained as part of the Spitzer First Look Survey.
To maximize the signal-to-noise of the measurement, only the central
two pixel radius of the PSF was used to scale the PSF and to derive 
our flux estimates.  
For MIPS 70$\mu$m, we used the raw data together with the MIPS-team DAT software
package \citep{gordon05} to produce final, mosaicked images - see
\citet{kim05} for details.
We do not detect any of our targets with confidence at 70$\mu$m.
We therefore followed the standard FEPS practice of deriving upper 
limits (or measured fluxes) from the MIPS 70$\mu$m images using 
aperture photometry, with a three pixel radius aperture.  

The Spitzer data for the Pleiades FEPS targets are provided
in Table 2.   The data for IRAC and MIPS24 are given in magnitudes;
the data for IRS24, IRS33 and MIPS70 are given in mJy.  
For pure photospheres, G dwarfs are expected to have
essentially constant magnitude at Spitzer wavelengths - hence by
presenting the data in magnitudes, one can very quickly scan the table and 
ascertain that most of the Pleiades stars have no significant IR
excess through 24$\mu$m.  The IRS and MIPS70 data are given in mJy
because there is no zero point calibration available to define
a magnitude system for IRS and because some of the IRS and MIPS
in-band fluxes are negative (consistent with non-detection 
within the errors), and hence cannot be straightforwardly
represented as magnitudes.
Flux zero points for the IRAC and MIPS24 bandpasses
are provided at the bottom of the table.  Based on a combination of
the internal uncertainty estimates provided by our software routines,
the consistency checks we have made (some of which are described
below), and other tests we have made using data for the older
stars in the FEPS sample, we believe that the typical 1$\sigma$\ 
statistical random
uncertainties in our flux estimates are 2\% for IRAC, 50 $\mu$Jy
for MIPS24 (i.e. also about 2\%), 0.4 mJy for IRS24, 0.45 mJy IRS33,
and about 7 mJy for MIPS70.  Note, however, that these estimated
1$\sigma$\ uncertainties are specific to the manner in which the
observations were obtained; for each instrument, our observations
were obtained in a very homogenous fashion and during a single
campaign (for MIPS, during a single, contiguous 10 hour period).
This likely results in relative uncertainties
which are smaller than is typically true for ``random" data sets.

%Figure \ref{fig:typSED} illustrates
%the quality of the data and our fitting procedure for a ``typical"
%Pleiades star, HII1182.  The solid black line in the figure is a
%best-fit Kurucz model whose parameters were determined from a
%least-squares fit to the optical and near-IR photometry.
%To normalize the Kurucz spectrum, we have UBVRIJHK
%photometry available for this and most
%other Pleiades members.   Because of relatively large variability
%and potential problems with absolute calibration and fidelity of the
%Kurucz models in the far blue, we did not use the U band photometry
%in our fits.  

As the simplest means to assess whether any of the Pleiades FEPS
targets shows evidence of an excess in the mid-IR,
Figure \ref{fig:Kminus24} shows the distribution of the K minus
MIPS24 magnitude difference for our sample.
The fact that most of the stars are distributed within 5\% of
$\Delta$mag = 0.00 in this diagram indicates that
(a) most of
the Pleiades stars do not have excesses and (b) the use of the
2.2 $\mu$m flux plus a Rayleigh-Jeans approximation to predict
the photospheric flux at 24 $\mu$m works quite well.   The five
stars with apparent excesses exceeding 0.1 mag are potential
debris disk detections, which we will examine in more detail
in the next section.

\section{Initial Analysis: Do the Pleiades G Dwarfs have Detected
Debris Disks?}
\label{sec:analysis}

Because the Spitzer data cover a broad wavelength regime, we are able
to detect dust with a wide range in temperature.  The excesses
we expect might be present are small, however, and an important
facet of our analysis is an investigation to verify that any
excesses we detect are real.  If the Pleiades stars do indeed have
mid-IR excesses, we need to determine how well our measurements
constrain parameters describing the circumstellar dust around these
stars, and hence our ability to provide quantitative constraints
on debris disk evolution for stars of mass similar to the Sun.
We address these questions in the next several sections.

\subsection{Do the Pleiades G Dwarfs Have Excesses at IRAC Wavelengths?}
\label{sec:IRAC_excess}

Based on standard, pre-launch model predictions,
we did not expect our 100 Myr old Pleiades G dwarfs to have a sufficient
amount of warm (T $>$ 300 K) circumstellar dust to yield detectable excesses at
IRAC wavelengths.  Nevertheless, we believe it is worthwhile to check
our IRAC data carefully for this possibility.  First, IRAC provides
the most sensitive and potentially the most stable camera at these
wavelengths to date, and hence it may be possible to detect excesses
from zodiacal-like dust
that would have escaped previous surveys.  Second, a few of the Pleiades
A dwarfs show excesses at 8 $\mu$m and at longer wavelengths \citep{stauffer05,
gorlova05}.   We believe that at least in most cases 
those excesses are not due to circumstellar dust disks but are instead
the product of interstellar dust (mostly emission from 
polycyclic aromatic hydrocarbons - PAH's) from the colliding molecular
cloud being ``lit up" by close passage by the Pleiades stars.  
Detection of an 8$\mu$m excess in any of our Pleiades G dwarfs might
be a signature of this phenomenon rather than of a debris disk, and
complicate interpretation of any excess we detect at longer wavelengths.

In practice, we do not see any evidence for excesses at IRAC wavelengths.
Figure \ref{fig:KminusCh2} illustrates this via a plot of the K - [4.5]
color for our Pleiades stars, while Figure \ref{fig:Ch2minusCh4} shows
the [4.5] - [8.0] color for our targets.   We expect these colors to
be essentially zero for G dwarfs, and to within 0.01 magnitude the mean
colors are zero for our sample (the mean color is -0.008 in both cases).   
The absolute calibration of the IRAC
photometry is not sufficiently well known to consider these mean offsets
as real.  The dispersion about the K - [4.5] mean color is 0.023 mag,
which is barely larger than the expected 0.02 mag uncertainty for
the 2MASS K magnitudes.   The dispersion about the [4.5] - [8.0] mean color
is only 0.014 magnitudes, indicative of both a lack of warm dust in the
Pleiades G dwarfs and the very good differential photometry possible
with IRAC.

\subsection{Do the Pleiades G Dwarfs Have Excesses at $\lambda$ $>$ 10$\mu$m?}
\label{sec:MIPS_excess}

\begin{figure} %
     \includegraphics[angle=00,totalheight=6.5in]{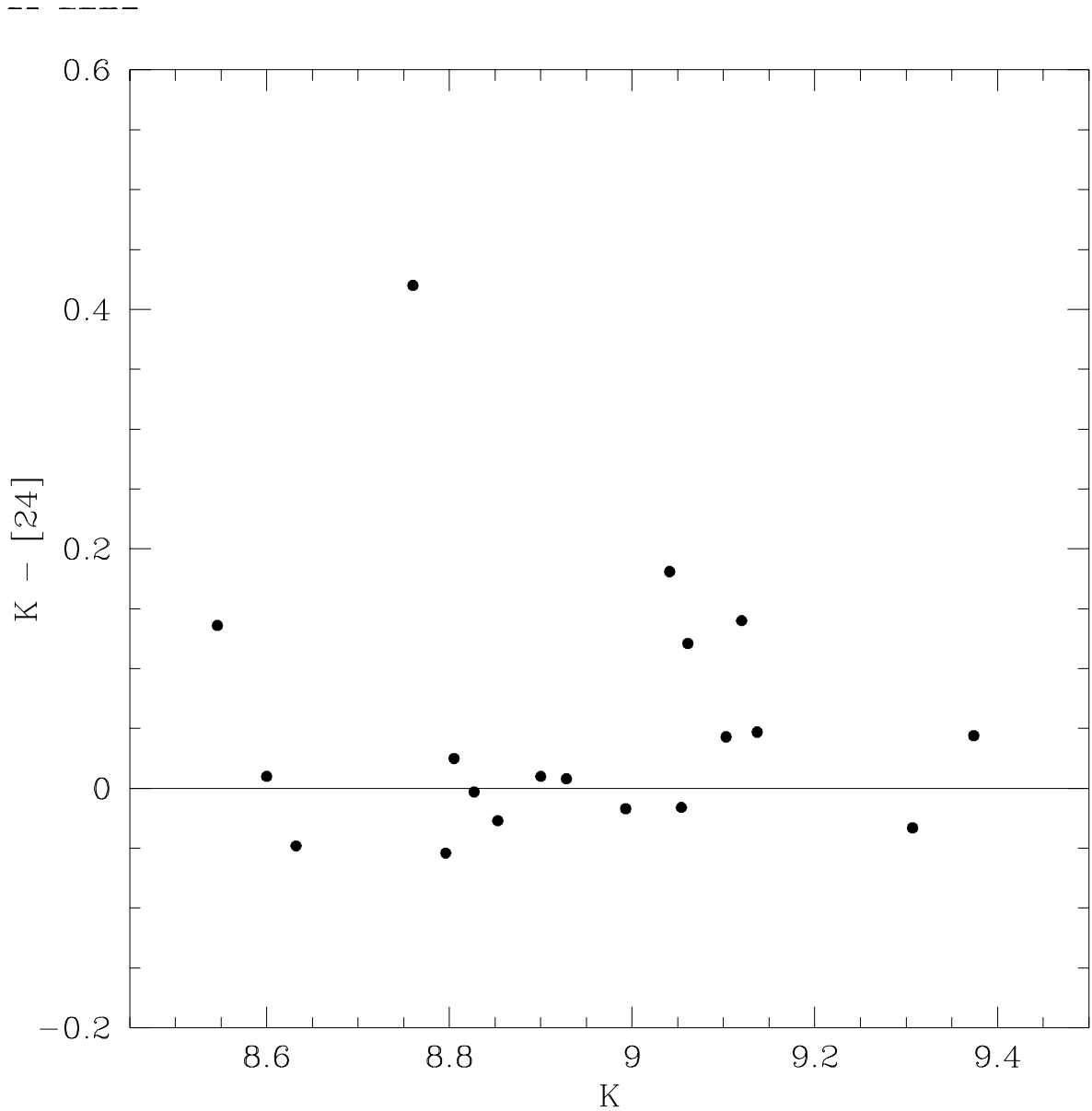}
    \caption{
        Plot of the difference between the  K band magnitude and the
        MIPS 24 $\mu$m magnitude for the Pleiades FEPS stars as a function
        of K-band magnitude.   For solar-type stars, the difference should
        be approximately zero for a pure photosphere.   The five stars with
        K - [24] magnitude differences greater than 0.1 are stars with
        possible excesses due to circumstellar debris disks.  The horizontal line
        simply marks equality between the two magnitudes.  The stars with the
        largest K - [24] colors are HII1101 (K - [24] = 0.420 and HII514
        (K - [24] = 0.18).
    \label{fig:Kminus24}
    }
\end{figure}

\begin{figure} %
     \includegraphics[angle=00,totalheight=6.5in]{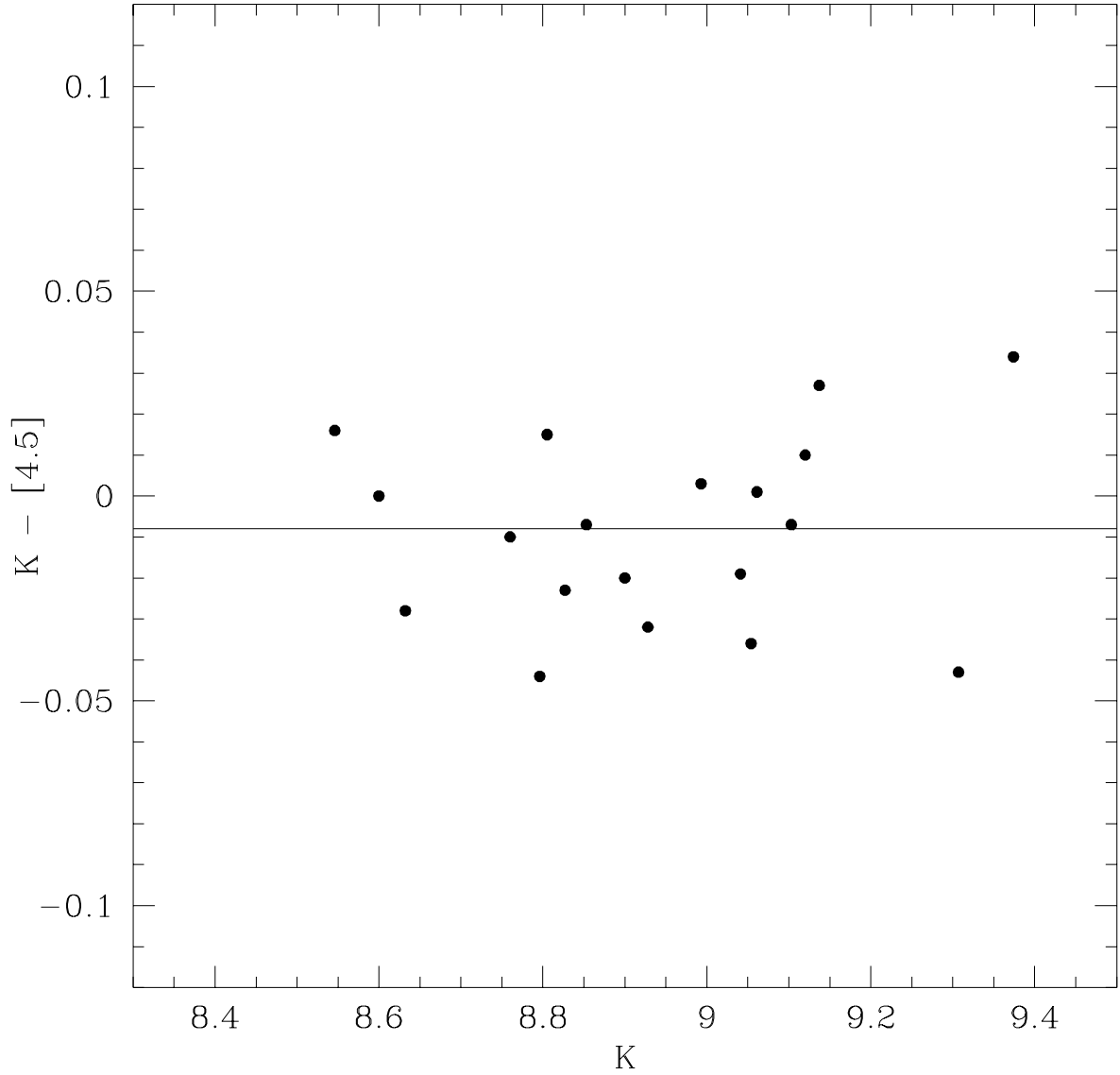}
    \caption{
        Plot of the difference between the K band magnitude and the
        4.5 $\mu$m magnitude for the Pleiades FEPS stars.   For solar-type
        stars, the difference should
        be approximately zero for a pure photosphere.  Because the Pleiades
        stars are all likely low amplitude photometric variables, the dispersion
        about zero represents an upper limit to the relative photometric accuracies of
        the K and IRAC Ch. 2 data.  The horizontal line indicates the mean
        K - [4.5] color for the Pleiades stars.
    \label{fig:KminusCh2}
    }
\end{figure}

\begin{figure} %
     \includegraphics[angle=00,totalheight=6.5in]{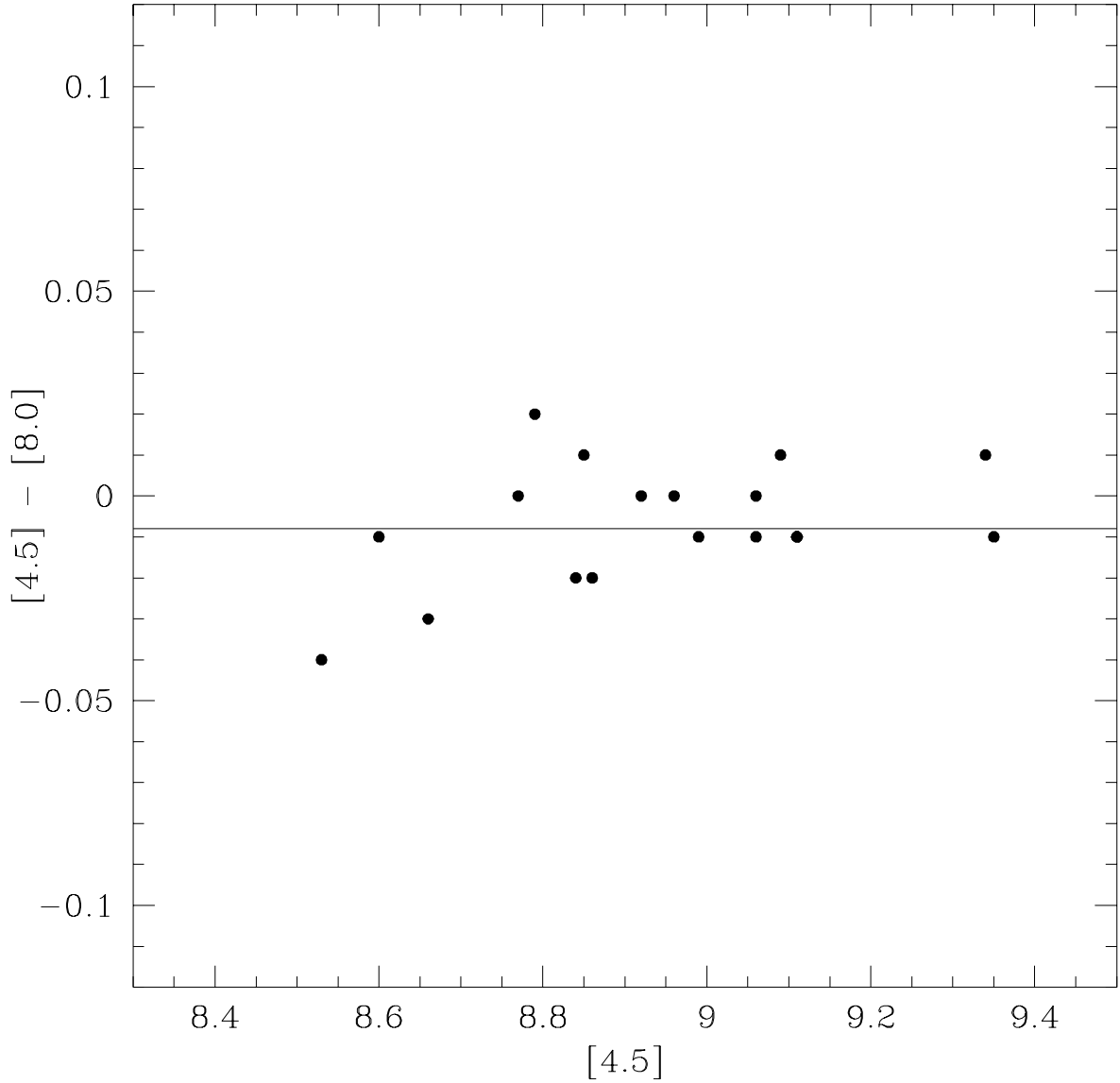}
    \caption{
        Plot of the difference between the Ch.2 (4.5$\mu$m) magnitude and 
        the Ch. 4 (8.0$\mu$m) magnitude for the Pleiades FEPS stars.   
        For solar-type stars, the difference should
        be approximately zero for a pure photosphere.  These observations are
        obtained simultaneously, so they are unaffected by the likely
        intrinsic variability of the Pleiades stars.  The horizontal line
        indicates the mean [4.5] - [8.0] color for the Pleiades stars.
    \label{fig:Ch2minusCh4}
    }
\end{figure}

In Figure \ref{fig:Kminus24}, we used the K - 24$\mu$m color as a simple
means to identify stars with mid-IR excesses.  A potentially more sensitive test
can be made by fitting a Kurucz model photosphere to the existing optical
and near-IR photometry for each star, and then subtracting that photosphere
from the Spitzer photometry.
Table 3 provides our predicted model in-band fluxes
for MIPS24, IRS33 and MIPS70.  The IRS24 model fluxes are 
identical to those for MIPS24.  The formal 1$\sigma$\ 
uncertainties for each of the predicted model in-band fluxes ranges
from about 2.5\% to 5\%, with typical values of order 4\%.

\begin{figure} %
     \includegraphics[angle=00,totalheight=6.5in]{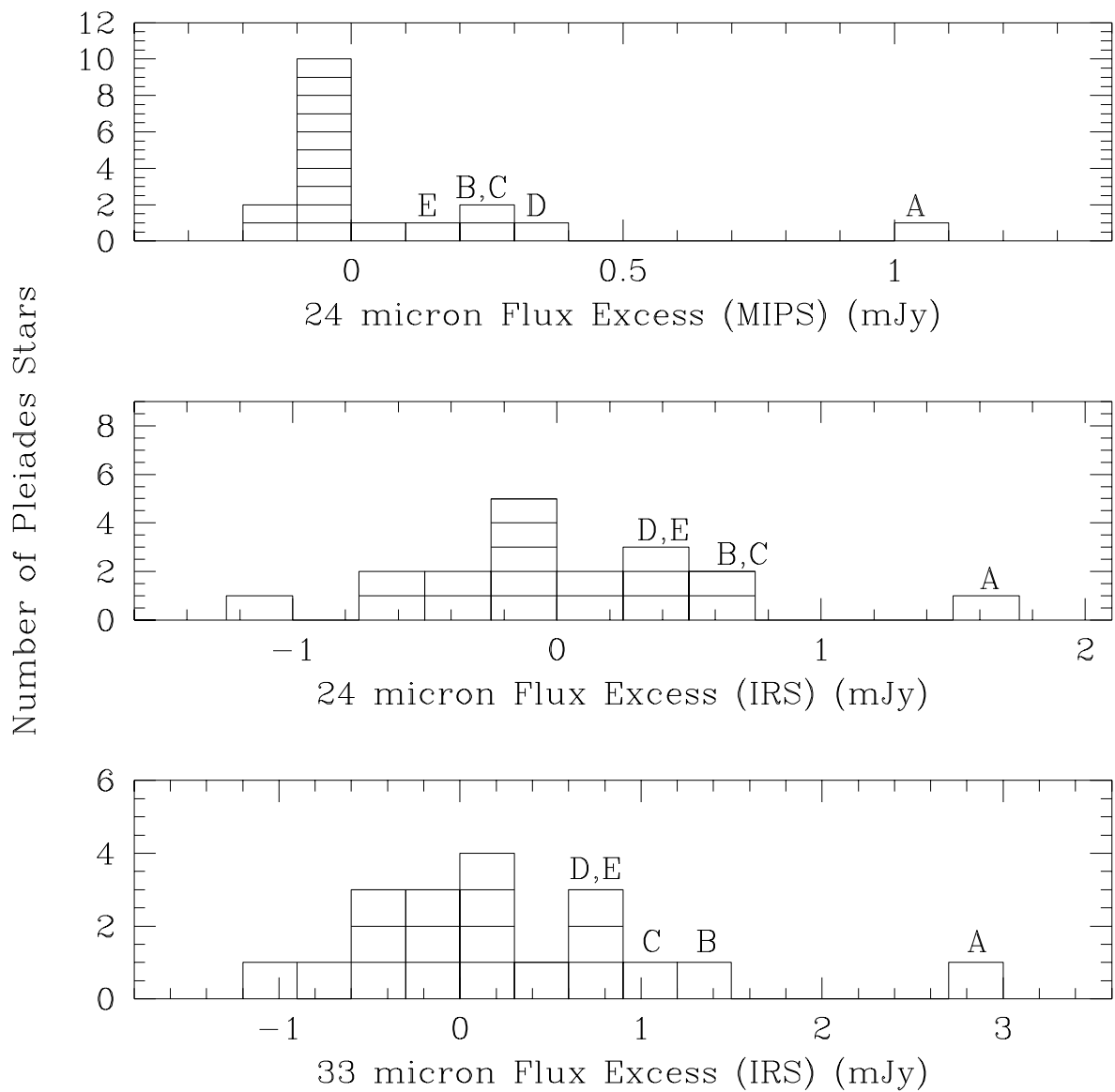}
    \caption{
       Differences between our measured in-band fluxes for MIPS24, IRS24
       and IRS33 and predictions from best-fit Kurucz models which have been scaled
       to the published optical and near-IR photometry for our targets.
       The labelled histogram boxes mark the positions of HII1101 (A), HII152 (B),
       HII514 (C), HII1200 (D) and HII250 (E).  See text for a discussion.
    \label{fig:plefeps.fig6}
    }
\end{figure}

Figure \ref{fig:plefeps.fig6} shows histograms of the flux excesses
(observed flux minus photospheric estimate) for
MIPS24, IRS24 and IRS33.  Each histogram is characterized by a 
set of stars whose calculated excesses scatter about zero and a small set of
stars populating a tail to positive excess, representing possible
debris disk detections.  Is this tail of IR-bright stars simply 
ascribable to random errors, or is it indicative of a population of
stars with real excesses?  Evidence for the latter conclusion is
provided by marking the location of five particular stars in the three
histograms:  HII1101 (A), HII152 (B), HII514 (C), HII1200 (D),
and HII250 (E).   With only one exception, these five stars are the
highest excess objects in each histogram, usually in the same order.
They are also the same five stars with K - MIPS24 colors greater
than 0.1 mag highlighted in Figure \ref{fig:Kminus24}.  For
MIPS24, all five have excesses at 4$\sigma$ or greater above
the photospheric prediction.
We conclude that there is real excess flux in our aperture
towards these stars.   However, the excess flux is not necessarily
associated with our Pleiades stars.

\begin{figure} %
     \includegraphics[angle=00,width=6.5in]{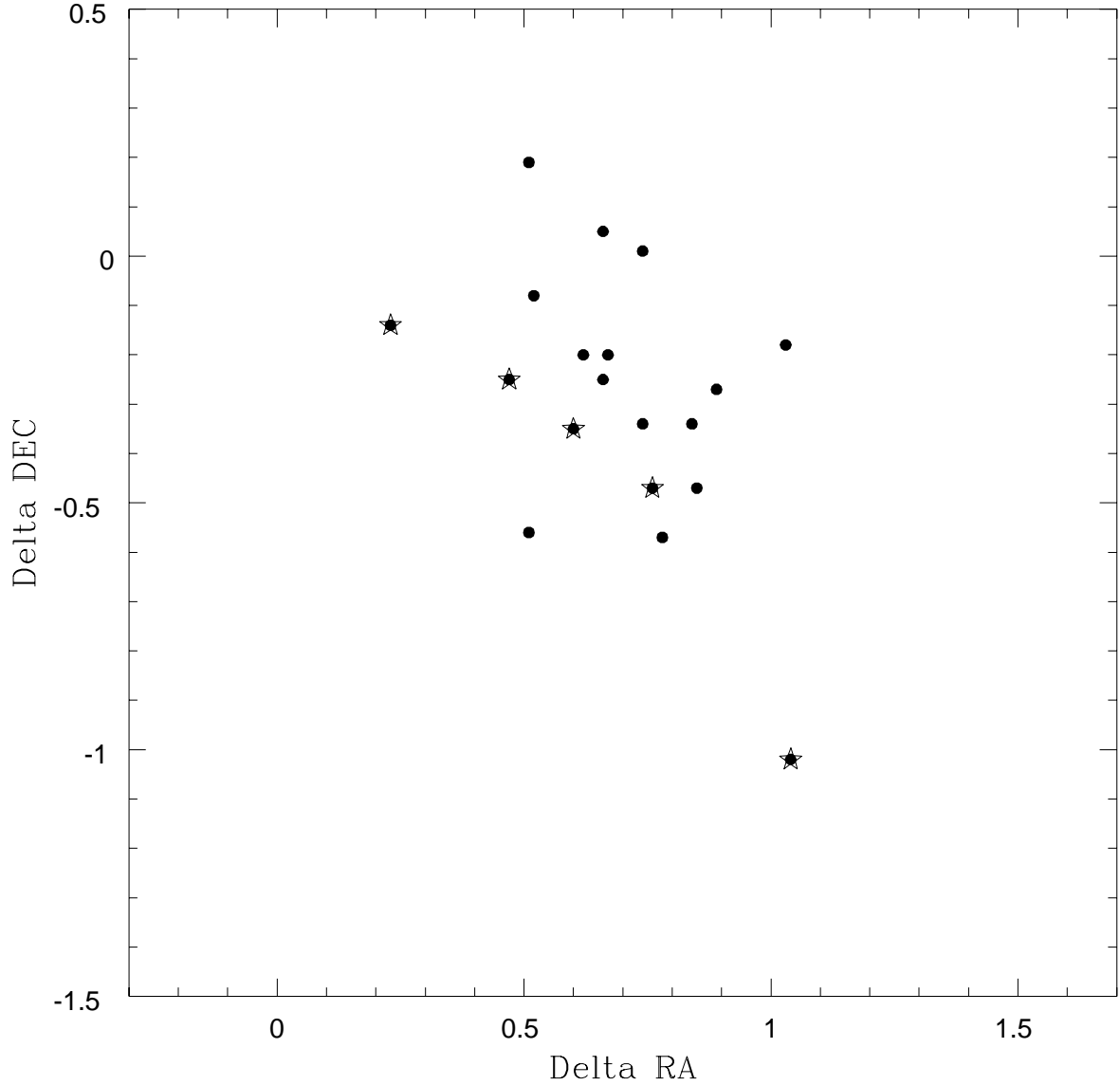}
    \caption{
         Offset between the 2MASS position of the Pleiades FEPS stars and
           the measured centroid positions from the MIPS24 images.  The 
           star symbols mark the five stars with the largest 24$\mu$m
           excesses (HII152, 250, 514, 1101 and 1200); it is HII152 that is at
           $\Delta$RA $\sim$1.0, $\Delta$DEC $\sim$ -1.0.  
           The large positional offset relative to the mean for HII 152
           is evidence that its 24$\mu$m excess is probably due to
           contamination of our flux measurement by local cirrus or a 
           background AGN near the line of sight.
    \label{fig:coords24}
    }
\end{figure}

The long-wavelength excesses could, in principle, be due to chance
alignments of our targets with uncatalogued asteroids.  However, because
the IRS and MIPS data were obtained at significantly different dates
yet the same stars show up with possible excesses, we believe that 
asteroids are not likely to be the cause of the excesses.
Instead, the most likely ``contaminant" 
to the 24 $\mu$m in-band fluxes is confusion
from random line-of-sight positional overlap with (a) distant
optically faint but IR-bright AGN or (b) inhomogeneities (cloudlets?
filaments?) in the Pleiades nebular emission.  One way to determine
if contamination is a problem is to look for positional offsets between the
measured MIPS24 positions of our targets and the input positions
for each AOR (which come from 2MASS for our targets).  This
test is only possible because of the excellent pointing accuracy
and stability of Spitzer; IRAC observations indicate that the
``blind-offset" positional accuracy for Spitzer is generally about
0.25 arcsec, with an RMS of similar magnitude.  Figure \ref{fig:coords24}
shows the positional offset between the 2MASS and MIPS24 centroids
for the Pleiades stars, with the five potential excess sources
highlighted.  One excess star - HII152 - is clearly an outlier.
If we exclude that star, the mean positional offsets are
$\Delta$RA = 0.62 arcsec and $\Delta$DEC = -0.24 arcsec; the RMS
of the radial positional error
is 0.27 arcsec - in excellent agreement with the expected pointing
accuracy of Spitzer and suggestive that other sources of error
(e.g. AGN contamination) do not significantly affect the Pleiades
measurements in general.   
Of the possible excess sources, HII152 has a $>$3 sigma offset from
its expected position, suggesting that its 24 $\mu$m flux is probably
contaminated. HII250 is the next most discrepant point (at $\delta$Dec
$\sim$ -0.13, $\delta$RA $\sim$ 0.23).  The
other three stars from Figure \ref{fig:coords24} 
(HII1101, 514 and 1200) have positional offsets
relative to the mean of the distribution $<$ 0.3 arcseconds,
consistent with no contamination.

Another means to assess whether an additional object near the line
of sight to our Pleiades stars is affecting our MIPS24 photometry
is to do PSF subtractions.  We have done that for our five stars
with the largest possible 24 $\mu$m excesses.
We performed two experiments: (a) subtract a 24 $\mu$m
PSF scaled to the measured flux from the mosaiced 24 $\mu$m (post BCD)
image; and (b) subtract a 24 $\mu$m PSF scaled to the predicted
flux from the Kurucz model.  Only HII152 showed any evidence for
a distorted, inner PSF in these experiments.  However, both HII152
and HII250 showed evidence of an extended, very low level pedestal
of emission (above the average sky background) extending 5-10 pixels
from the star.  Because this is too large an angular extent to be
indicative of a dusty debris disk centrally heated by our target
star, we assume it is indicative of a local condensation of the
interstellar nebulosity in the Pleiades, projected towards the line of sight.
To help further elucidate
the nature of the possible 24$\mu$m excesses,
we also determined aperture growth curves (for apertures from
2 to 7 pixel radius) for the five possible excess stars and for a sample
of the stars with no apparent excess.  Most of the stars - including HII1101, 
the star with the largest excess - showed growth curves consistent
with expectation for the STINYTIM model PSF.   Both HII152 and HII250
show growth curves that are broader than that expected for pure
point sources, consistent with extended local nebular emission in the
source aperture above
the background estimated from our 10 to 20 pixel radius sky annulus.
For both these stars, the growth curves are well matched by a model
consisting of a standard PSF
plus a constant excess sky out to 7 pixel radius.  However, the predicted
effect on our PSF-fitting flux of this local excess nebular emission
is only about 1/3 of the excess that we measured (Table 2).  Either the
local nebular emission must have a sharp peak at the location of 
HII152 and HII250, or other effects may also be influencing their 24$\mu$m
flux.  HII514 and HII1200 show slight irregularities in their growth curves -
due to a faint ``companion" at 4.5 pix distance for HII514 and probably
due to local cirrus for HII1200.
We believe the PSF flux we report for them is correct, but with less
certainty than for most of our stars.

Finally, it is possible to estimate the number of expected cases
where a background, IR bright galaxy would affect our MIPS Pleiades
photometry based on Spitzer 24$\mu$m source count statistics
\citep{papovich04}.   Figure 2 of that paper shows cumulative
24$\mu$m source count data derived from several different high
latitude fields; that plot indicates expected cumulative source
counts of order 10$^6$ per steradian for S$_{\nu}$ $>$ 1 mJy
and 10$^7$ per steradian for S$_{\nu}$ $>$ 0.3 mJy.  First,
consider HII1101 - our star with the largest excess at 24$\mu$m,
about 1 mJy.  The cumulative counts for that flux level correspond
to about one source per 40,000 square arcseconds.  In order for
a 1 mJy source to displace the centroid position of HII1101 by
less than 0.3 arcseconds, it would have to be located less than
an arcsecond from HII1101 (given HII1101's photospheric flux at
24$\mu$m of 2.2 mJy and assuming our measured excess is due to
a background galaxy).  The a posteriori probability of a background
galaxy with 24$\mu$m flux $\geq$ 1 mJy being this close to the line of
sight to HII1101 is therefore less than 1 in 10,000.  
For our sample of 20 targets, there is then a $\sim$1 in 500 chance
that one of them would have had this happen, supporting our belief
that the excess for HII1101 has a high probability of being real.
Would AGN contamination provide a plausible explanation for the large
positional error and 24 $\mu$m excess we found for HII152?  Compared 
to HII1101, the 24$\mu$m excess for HII152 is smaller, of order 0.3 mJy,
and the derived 24$\mu$m positional error is larger, of order
0.9 arcseconds.  Using the 10$^7$ sources per steradian number
appropriate for this flux level, we derive a probability of a
background source being this close to the line of sight of a given Pleiades
star of order 1 in 40.  Therefore, for our sample of 20, we should
have expected a 50\% probability that one of them would have had
the properties we found for HII152, supporting AGN contamination as
a plausible explanation for the 24$\mu$m excess of this star.  With
fairly similar properties, HII250 is also a good candidate for AGN
contamination.   Lastly, we consider the other two possible excess
sources - HII514 and HII1200 (MIPS24 flux excess
of about 0.3 mJy, MIPS24 positional offset $<$0.3 arcsec, and
photospheric flux of $\sim$2 mJy).  We derive about a 1 in 400 chance
of this alignment for a single target, or about a 5\% probability
that one of our sample of 20 would have such an alignment.   
We conclude that the excesses for HII514 and HII1200 are 
plausibly real, but with significantly less confidence than for
HII1101.   In summary, our best judgment for the five stars with 
24 $\mu$m excesses are:
\vskip0.1truein
\noindent
HII152:  very likely contaminated by cirrus or background AGN.   Not a debris disk.
\vskip0.01truein
\noindent
HII250:  probably contaminated by cirrus or background AGN.
\vskip0.01truein
\noindent
HII514:  possible debris disk.
\vskip0.01truein
\noindent
HII1200:  possible debris disk.
\vskip0.01truein
\noindent
HII1101:  probable debris disk.

\section{Discussion}
\label{sec:discussion}

\subsection{Physical Interpretation of the Excesses and Upper Limits }
\label{sec:physical}

\begin{figure}
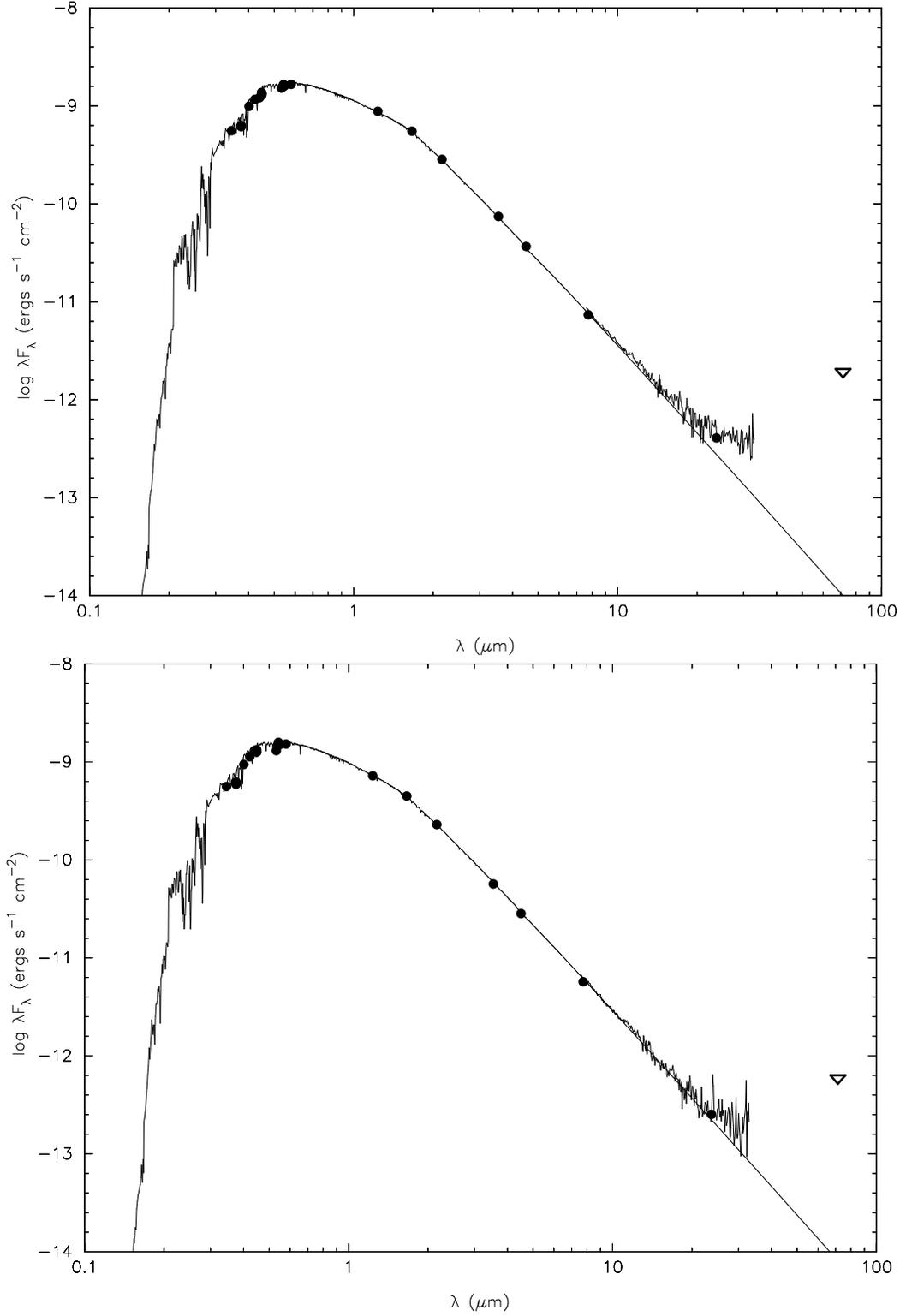
 %
     \includegraphics[angle=00,width=4.0in]{f8a.ps}
     \vspace{2.5in}
     \includegraphics[angle=00,width=4.0in]{f8b.ps}
    \caption{
        Spectral energy distributions and best-fit Kurucz models for the
        two Pleiades stars with the best evidence for mid-IR excesses 
        (top: HII1101; bottom: HII514).
    \label{fig:excessSED}
    }
\end{figure}

Figure \ref{fig:excessSED} shows the SED's for the Pleiades
FEPS target with the best evidence for mid-IR excess (HII1101)
and one of the two stars with possible excesses (HII514).
The solid black line in the figure is a
best-fit Kurucz model whose parameters were determined from a
least-squares fit to the optical and near-IR photometry.
We used published BVRIJHK photometry to normalize the Kurucz 
model spectra. 

For the purposes of the physical model fits, we assume a Pleiades
distance of 133 pc, which is the average of several recent
determinations using astrometric observations of binary
star Pleiades members \citep{pan04,zwahlen04,munari04},
direct trigonometric parallax using HST \citep{soderblom05},
and metallicity-corrected main-sequence fitting \citep{percival04}.
In order to convert our observables to something more physical which
can be usefully compared, for example, to our solar system's Kuiper
belt or to debris disks of other ages, it is useful to calculate a few
physical parameter constraints for our Pleiades sample based on a
specific, simple model.  We describe this model in detail below.  This
particular model provides minimum dust grain masses which are consistent
with our Spitzer data.

We fit the excess SED (observations minus estimated photosphere) with
a simple blackbody grain model based on the color temperatures 
(T$_c$) calculated
from the Planck formula and the ratio of
excess fluxes measured in the MIPS bandpasses and our IRS33 bandpass.
Increasing excess flux densities between $<$24$\mu$m and 33$\mu$m indicate that we
are observing the Wien side of the SED, with the color temperature 
corresponding
approximately to the maximum dust temperature in the disk, and hence 
the minimum
distance from the star of the emitting material.  The relationship 
between grain
temperature, position and primary star luminosity (Backman \& Paresce 1993)
for ``blackbody" grains
that absorb and emit radiation efficiently at all observed wavelengths is: 
\begin{equation}
%R$_{in}$ = (278K * (L$_{star}$/L$_{sun}$)$^{0.25}$/T$_c$)$^2$ 
R_{in} = (278K * (L_{star}/L_{sun})^{0.25}/T_c)^2 \end{equation}

The derived dust temperature T$_c$, observed excess flux densities, and cluster
distance can then be used with the Planck and Stefan-Boltzmann equations
to calculate a total grain cross-sectional area and dust luminosity.
An estimate for the total dust mass can then be derived from the
cross-sectional area by assuming a material density and a characteristic
grain radius consistent with the assumption of blackbody emission to the
maximum wavelength of the observed excess \citep{kim05}.
The total radiating dust mass estimates derived are minimum values because:
a) they are derived from the dust maximum temperature whereas a broader radial
range including lower temperature dust would have more mass for a given
luminosity, and b) they assume a single dust grain size, whereas dust
with a size distribution (e.g. a ``standard" MRN model
in which n $\propto$ a$^{-3.5}$) has most of the luminosity arising from the
smallest grains but most of the mass residing in the largest grains.

For HII1101, following the process described above, the inferred
24/33 micron color temperature of the excess is 84K.  Using
this color temperature, we derive R$_{in}$ = 12.6 AU.  For our
measured MIPS 24$\mu$m excess of 1.09 mJy, we obtain a dust
solid angle of 4.8$\times$10$^{-16}$ steradians,  and with 133 pc
for the distance to the Pleiades we then derive the total dust
cross-sectional area of A = 0.34 AU$^{2}$.  With a generic silicate density of
2.5 g cm$^{-3}$, and a dust radius of 10 microns, this yields a lower limit
to the dust mass of Log(M$_{dust}$/M$_{Earth}$) = -4.4, and a
fractional dust luminosity- Log(L$_{dust}$/L$_*$) = -3.8.

We can place an upper limit to the debris disk mass for the $\sim$17
Pleiades solar-mass stars for which we do not detect mid-IR excesses.
Using the same formalism as for HII1101, the grain temperature derived for HII1101,
and the 3 $\sigma$\ RMS upper limits for the non-detected stars
- 150$\mu$Jy and 21 mJy, respectively, for MIPS24 and MIPS70 - we derive
estimated 3$\sigma$ upper limits for dust mass of
   Log(M$_{dust}$/M$_{Earth}$) = -5.2 and -3.8.   Because the fluxes
radiated at the shorter wavelengths are more dependent on the details of
the dust grain size distribution and composition and the radial distribution
of dust, the dust mass upper limit for MIPS70 is less model dependent 
than for MIPS24.

The SED for HII1101 implies that there is less dust radiating at temperatures
above 84K than at or below 84K.  A rough estimate of an upper limit 
to the amount
of material with T $>$ 84K (r $<$ R$_{in}$ $\sim$ 12.6 AU) can be 
calculated
such that emission from the warm dust would not exceed 1$\sigma$ above the
observed SED already fit by the photospheric emission plus
blackbody dust excess model described above.
We assume a distribution of dust extending from 12.6 AU to silicate
sublimation at T $\sim$ 1500K (r $\sim$ 0.04 AU) with a radially constant
surface density consistent with dynamics controlled by Poynting-Robertson
(PR) drag.  Under these assumptions the upper limit on dust warmer than 84K is
Log(M$_{dust}$/M$_{Earth}$) $<$\ -5.7.  There is evidently a central zone
relatively depleted of dust inside the zone in which dust is detected.
The black-body
grain model described above would require something to prevent the
dust grains generated in a planetesimal belt beyond 13 AU from
spiraling into the inner dust due to PR drag. While other models
cannot be ruled out, it is interesting to note that the presence of
giant planet inside of 13 AU could be responsible for the inner hole
\citep{kim05}.

\subsection{Implications for Debris Disk Evolutionary Timescales}
\label{sec:implications}

How do our observed properties of solar-mass stars in the Pleiades
compare to what we believe the debris disk for an average star like the Sun
should look like at an age of 100 Myr?   We address this question
by using a simplified evolutionary model of our solar system's
Kuiper Belt (KB) \citep{kim05,backman05}.  The model
combines collisional evolution of the KB planetesimal population
with quasi-equilibrium dust production and dust spatial re-distribution
\citep{backman95}.  A parallel model of the evolution
of the asteroid belt and its associated zodiacal dust cloud predicts
negligible far-IR contribution from the inner solar
system relative to the outer solar system at the ages of the Pleiades
stars.

The model includes the effects of predicted much stronger winds
from young low mass stars.  In particular, at Pleiades
age we assume a wind mass loss rate 1000 times the current solar
wind, following \citep{wood02}.  The effect of this wind is to add an
additional scouring mechanism acting preferentially to remove small grains,
hence primarily affecting emission at shorter wavelengths.  For
Pleiades age, a model with 1000 times solar wind mass loss rate
shows a 60\% decrease in the 24$\mu$m flux density but only a 25\%
decrease in the predicted 70$\mu$m flux density relative to a model
with wind like the present-day Sun's.

In order to be generic, the model does not include some effects which
could significantly affect the debris disk characteristics at Pleiades
age but which are dependent on dynamical evolutionary events which
are neither deterministic nor easily incorporated into a simple
model like ours.   In particular, our model
implicitly assumes that the orbits of
the major planets did not change significantly between Pleiades age and the
present.  Some current research
indicates that considerable orbital migration may have occurred in our
solar system after an age of 100 Myr \citep{levison04,gomes05},
including events that could have temporarily greatly enhanced or 
reduced the IR luminosity of our system.
Our model also does not account for individual rare large planetesimal
collisions temporarily injecting large amounts of dust
which would have been more common in earlier eras.

With the above caveats, the predicted flux densities 
for 133pc,  100 Myr and for a solar-mass
star (and hence
for a luminosity 80\% that of the present-day Sun) are 0.4 mJy at
24$\mu$m and 12mJy at 70$\mu$m.  Those numbers bracket our 3$\sigma$\
upper limits for the non-detected Pleiades G dwarfs at those two wavelengths.
Each individual Pleiades star is therefore roughly
consistent with the prediction
of the heuristic model.  However, as an ensemble, the mean observed
24$\mu$m and 70$\mu$m excess fluxes for the Pleiades G dwarfs are
below the predictions of the model.
We note that if one simply put the Sun and {\it present-day}
solar system dust at Pleiades distance,
the approximate 24 and 70$\mu$m flux dust excesses we would observe
would be of order a few $\mu$Jy at 24$\mu$m and of order 50$\mu$Jy at 70$\mu$m,
well below what is detectable with Spitzer.

\subsection{Correlations with Other Stellar Properties}
 
With only one to possibly three stars with detected excesses, our statistical base
is too small to draw any definitive conclusions concerning the correlation
between IR excess and other stellar properties.  Hence we cannot
yet address with any certainty the questions raised in the
introduction concerning possible correlations between debris disks
and stellar mass, rotation, and xray luminosity .  In any event,
our probable excess star (HII1101) and our two possible excess
stars (HII514 and HII1200) are fairly unremarkable
in most of their properties.  They are not known to be binaries, but
that is also true of most of our targets.  Their projected rotational
velocities are average for our FEPS Pleiades sample (see Table 1),
and average for stars of their age.  Only three of
our Pleiades stars have rotational velocities more than 20 \kms;
those three stars do not have excesses, but $>$80\% of our stars
do not have excesses.  HII1200 is the most massive star 
in our sample, HII1101 is the fourth or fifth most massive star, while
HII514 is average in terms of its inferred mass.  This is perhaps slight
evidence in favor of the trend noted in the introduction in other datasets
for the stars with detected debris disks
to be higher mass.  There is a weak correlation in our data between xray 
activity (Log(L$_x$/L$_{bol}$) in Table 1) and debris disk detection in 
the same sense as seen by \citet{chen05} since our Pleiades sample 
includes a half dozen stars with approximately saturated xray 
emission (Log(L$_x$/L$_{bol}$) $\sim$ -3) and none of them have
detected IR excesses, while our three detected sources have
Log(L$_x$/L$_{bol}$) $\sim$ -4.2 or less.  The links between stellar
mass, xrays and mid-IR excess in our sample are the same as in the
\citet{chen05} sample, however, so selection effects may equally well
explain the apparent correlation between xray activity and mid-IR excess.
In order to draw any
statistically significant conclusions simply requires larger 
samples and, if possible, better 70$\mu$m limits.  The lack of 
any ${\it{strong}}$ correlations between
other observables and debris disk detection
in some sense favors a model where the detected stars just
happen to be the ones with the more recent, large collisional
events, as predicted by \citet{kenyon04}.  However, the alternative
hypothesis that there is simply a large range in debris disk masses
amongst the Pleiades stars and our detected objects are just those
with the most massive debris disks is also compatible with our data.

\section{Summary and Conclusions}

The Spitzer Space Telescope is providing the first statistical insight
into debris disks and debris disk evolution around solar-type stars, in 
which context for our own solar system may be found.  
From a study of twenty solar-mass Pleiades members from the FEPS Spitzer
Legacy program we have one probable and two possible detections
of stars with flux excesses indicative of
circumstellar dust.  Our results suggest that the frequency of stars 
with debris disks in excess of $\sim$0.3 mJy (24$\mu$m) at 100 Myr is 
thus $\sim$10\%.  We find that most of the sample have no detectable 
excess at Spitzer wavelengths with 3 sigma upper limits of 
0.15, 1.35, and 21 mJy, at 24, 33, and 70 $\mu$m respectively.  
Our results for the 100 Myr old
solar-type Pleiades members studied here are roughly consistent with (but
lower on average than) predictions
of a heuristic model \citep{backman05} for the evolution 
of the Sun's debris disk.

\acknowledgments

The authors thank Jean-Claude Mermilliod for information related to
HII1101.  We also thank the members of the MIPS DAT team (Karl Misselt,
James Muzerolle, Karl Gordon, Chad Engelbracht, Jane Morrison and Kate
Su) for assistance with the DAT software package, and the other
members of the FEPS team for their contributions to the overall 
FEPS program.  The FEPS team is pleased to acknowledge financial
support through NASA contracts 1224768, 1224634 and 1224566
administered by JPL.

%%
%% Bibliography
%%

%%
%% Tables
%%

\clearpage 
\begin{table}
\caption{Pleiades G and K Dwarf FEPS Targets}
\begin{tabular}{lcccccc}
\tableline
 Name &   V  & B-V &  vsini$^1$ & P  & Binarity? & Log(Lx/Lbol)   \\
      &      &     &  km-s$^{-1}$ & hours &      &             \\
\tableline
HII 120& 10.81 &0.70 &  9.4 &   -  & SB1?  & -  \\
HII 152& 10.73 &0.69 & 11.4 &   -   & N    & -  \\
HII 173& 10.87 &0.84 & 7.8/6.3 &  -  & SB2 & -4.20  \\
HII 174& 11.62 &0.85 & 90:  &   -   & N    & -3.05  \\
HII 250& 10.70 &0.68 &  6.4 & $<$24?  & N  & -4.06   \\
HII 314& 10.61 &0.65 & 41.9 & 35.4  & N    & -3.22 \\
HII 514& 10.72 &0.70 & 10.5 &   -   & N    & -4.22  \\
HII 1015&10.54 &0.65 &  9.6 &   -   & N    & -4.18 \\
HII 1101&10.26 &0.61 & 19.  &   -   & N    & -4.24  \\
HII 1182&10.46 &0.64 & 16.4 &   -  & VB    &  -  \\
HII 1200& 9.92 &0.54 & 13.7 &   -   & N    & $<$-4.96 \\
HII 1776&10.91 &0.72 & 10.1 &   -   & N    &  - \\
HII 2147&10.86 &0.81 & 6.9/10.8 & -  & SB2 & -2.94  \\    
HII 2278&10.90 &0.87 &  6.9 &   -  & VB    & -3.82  \\
HII 2506&10.25 &0.60 & 13.8 &   -   & N    &  -  \\
HII 2644&11.07 &0.75 &  4.3 &   -   & N    & -4.27  \\
HII 2786&10.30 &0.60 & 22.0 & 53.0  & N    & -4.30  \\
HII 2881&11.54 &0.96 & 10.5 &102.0 & VB    & -3.30  \\
HII 3097&10.94 &0.74 & 14.6 &  -   & SB1   &  - \\ 
HII 3179&10.05 &0.56 &  4.9 &  -    & N    & -4.42 \\
\tableline 
\end{tabular}
\vskip0.1truein
$^1$For SB2's, vsini's for both components are given.

\end{table}

\clearpage 
\begin{table}
\caption{Pleiades G and K Dwarf FEPS Targets}
\begin{tabular}{lcccccccc}
\tableline
      &          &   IRAC     &    IRAC    &    IRAC    &   MIPS    & IRS       &    IRS & MIPS$^1$    \\
 Name &     K    & 3.6 $\mu$m & 4.5 $\mu$m & 8.0 $\mu$m & 24 $\mu$m & 24 $\mu$m & 33 $\mu$m & 70 $\mu$m  \\
 Name &    mag   &    mag     &    mag     &   mag      &   mag     &   mJy     &    mJy    &    mJy     \\
\tableline
HII  120 & 9.103  &   9.075 &   9.113 &   9.119 &   9.06  & 1.41 & 0.48 &  $<$36.  \\ 
HII  152 & 9.12   &   9.069 &   9.108 &   9.119 &   8.98  & 2.19 & 2.09 &  $<$27.  \\
HII  173 & 8.827  &   8.797 &   8.852 &   8.844 &   8.83  & 2.03 & 1.24 &  $<$23.   \\
HII  174 & 9.374  &   9.292 &   9.344 &   9.330 &   9.33  & 0.66 & -0.07 & $<$17.   \\
HII  250 & 9.061  &   9.043 &   9.061 &   9.074 &   8.94  & 2.01 & 1.71 &  $<$15.  \\
HII  314 & 8.900  &   8.876 &   8.921 &   8.918 &   8.89  & 2.03 & 1.74 &  $<$23.  \\
HII  514 & 9.041  &   9.026 &   9.064 &   9.060 &   8.86  & 2.33 & 2.00 &  $<$15.  \\
HII 1015 & 8.993  &   8.958 &   8.985 &   8.997 &   9.01  & 0.56 & -0.05 & $<$30. \\
HII 1101 & 8.76   &   8.735 &   8.766 &   8.775 &   8.34  & 3.92 & 4.07 &  $<$24.  \\
HII 1182 & 8.928  &   8.917 &   8.955 &   8.963 &   8.92  & 1.85 & 1.10 &  $<$10.  \\
HII 1200 & 8.546  &   8.512 &   8.533 &   8.572 &   8.41  & 3.15 & 2.27 &  $<$43.  \\
HII 2147 & 8.60   &   8.555 &   8.596 &   8.605 &   8.59  & 3.05 & 1.50 &  $<$10.  \\
HII 2278 & 8.805  &   8.739 &   8.791 &   8.775 &   8.78  & 1.70 & 0.72 &  $<$23.  \\
HII 2506 & 8.796  &   8.805 &   8.844 &   8.862 &   8.85  & 1.93 & 1.25 &  $<$15.  \\
HII 2644 & 9.307  &   9.302 &   9.350 &   9.358 &   9.34  & 1.36 & 0.14 &  $<$15.  \\
HII 2786 & 8.853  &   8.832 &   8.858 &   8.880 &   8.88  & 1.90 & 0.92 &  $<$16.  \\
HII 2881 & 9.054  &   9.060 &   9.087 &   9.082 &   9.07  &  - &    -  &   $<$12. \\
HII 3097 & 9.137  &   9.070 &   9.113 &   9.119 &   9.09  & 1.60 & 0.62 &  $<$10.  \\
HII 3179 & 8.632  &   8.623 &   8.657 &   8.694 &   8.68  & 2.42 & 1.21 &  $<$10.  \\
\tableline 
\end{tabular}
\vskip0.1truein
IRAC Flux zero points: 277.5, 179.5, 63.1 Jy for 0 magnitude for Ch. 1, 2 and 4,
respectively \citep{fazio04}; the fluxes are not color-corrected.  One sigma,
statistical random uncertainties for all three IRAC bands are $\sim$ 2\%.
\vskip0.1truein
MIPS Flux zero point: 7.20 Jy for 0 magnitude for MIPS24 (from the MIPS
Data Handbook, v2.3 and
from data for FEPS G dwarfs older than 1 Gyr where no excess is expected).
Typical one sigma, statistical random uncertainties for the MIPS24 fluxes
are 50$\mu$Jy.
\vskip0.1truein
$^1$The indicated values are 3$\sigma$\ upper limits; none of the
stars are clear detections at MIPS70.
The average 1$\sigma$\ noise for these measurements is about 7 mJy, but
some of the stars have visible cirrus or faint objects near them and
their noise level is higher.  

\end{table}

\clearpage 
\begin{table}
\caption{Predicted Photospheric Fluxes for FEPS Targets}
\begin{tabular}{lcccccc}
\tableline
 Name & 24$\mu$m & 33$\mu$m & 70$\mu$m  \\
      &   mJy    &   mJy    &   mJy     \\
\tableline
HII 120 & 1.677 & 0.887 & 0.188 \\
HII 152 & 1.619 & 0.854 & 0.181 \\
HII 174 & 1.299 & 0.687 & 0.146 \\
HII 173 & 2.196 & 1.163 & 0.246 \\
HII 250 & 1.704 & 0.900 & 0.191 \\
HII 314 & 2.010 & 1.065 & 0.226 \\
HII 514 & 1.793 & 0.949 & 0.201 \\
HII 1015 & 1.846 & 0.975 & 0.206 \\
HII 1101 & 2.231 & 1.177 & 0.249 \\
HII 1182 & 1.939 & 1.024 & 0.217 \\
HII 1200 & 2.785 & 1.473 & 0.311 \\
HII 1776 & 1.591 & 0.842 & 0.178 \\
HII 2147 & 2.669 & 1.416 & 0.300 \\
HII 2278 & 2.286 & 1.213 & 0.257 \\
HII 2506 & 2.228 & 1.178 & 0.249 \\
HII 2644 & 1.385 & 0.732 & 0.155 \\
HII 2786 & 2.097 & 1.108 & 0.234 \\
HII 2881 & 1.755 & 0.929 & 0.197 \\
HII 3097 & 1.679 & 0.889 & 0.188 \\
HII 3179 & 2.559 & 1.353 & 0.286 \\
\tableline 
\end{tabular}

\end{table}

%%
%% Figures and Captions.
%%

\clearpage
\newpage

%\pagestyle{myheadings}
%\pagenumbering{arabic}
%\setcounter{page}{1}

%% 
%% Pleiades CMD Figure 1
%%

\clearpage
\newpage
%% \markright{Figure 1}

%% 
%% Histogram of CMD displacements
%%

\clearpage
\newpage
%% \markright{Figure 2}

%% 
%% Typical SED for Pleiades FEPS Target
%%

%\clearpage
%\newpage
%% \markright{Figure 3}
%
%\begin{figure} % 
%     \includegraphics[angle=00,totalheight=6.5in]{Figures/HII_1182.jmc.ps}
%    \caption{
%       Plot of ground-based and Spitzer data for HII1182, a typical Pleiades
%       FEPS target (i.e. one with no evidence of a significant mid-IR excess).
%       The curve is a Kurucz model fit to the optical and near-IR photometry.
%    \label{fig:typSED}
%    }
%\end{figure}

%%
%% Histogram of MIPS 24 micron flux excesses
%%

\clearpage
\newpage
%% \markright{Figure 3}

%%
%% K minus Ch. 2 color for Pleiades G dwarfs
%%

\clearpage
\newpage
%% \markright{Figure 4}

%%
%% Ch. 2 minus Ch. 4 color for Pleiades G dwarfs
%%

\clearpage
\newpage
%% \markright{Figure 5}

\clearpage
\newpage
%% \markright{Figure 6}

%%
%% MIPS 24 position error plot
%%

\clearpage
\newpage
%% \markright{Figure 7}

%%
%% SED's of Pleiades stars with excesses
%%

\clearpage
\newpage
%% \markright{Figure 8}

\end{document}